\documentclass[aps,prb,twocolumn,showpacs,floatfix]{revtex4}

\usepackage{graphicx}

\begin{document}

\title{Electronic structure and magnetic properties of graphitic ribbons}

\author{L. Pisani$^1$, J. A. Chan$^1$,  B. Montanari$^2$, and N. M. Harrison$^{1,3}$}

\affiliation{$^1$Department of Chemistry, Imperial College London, South Kensington campus, London SW7 2AZ, United Kingdom \\ $^2$CCLRC Rutherford Appleton Laboratory, Chilton, Didcot, Oxfordshire OX11 0QX, United Kingdom \\ $^3$CCLRC Daresbury Laboratory, Daresbury, Warrington WA4 4AD, United Kingdom}

\pacs{pacs}
\date{\today}

\begin{abstract}

First principles calculations are used to establish that the electronic
structure of graphene ribbons with zig-zag edges is unstable with respect
to magnetic polarisation of the edge states. The magnetic interaction between
edge states is found to be remarkably 
long ranged and intimately connected to the
electronic structure of the ribbon. Various treatments of electronic exchange
and correlation are used to examine the sensitivity of this result to details
of the electron-electron interactions and the qualitative features are found
to be independent of the details of the approximaton. The possibility of other
stablisation mechanisms, such as charge ordering and a Peierls
distortion, are explicitly considered and found to be unfavourable for
ribbons of reasonable width. These results have direct implications for the
control of the spin dependent conductance in graphitic nano-ribbons using
suitably modulated magnetic fields.

\vspace{1cm}

\footnotesize{}
 
\end{abstract}
\maketitle

\section{Introduction}
\label{intro}
 
Recently, room temperature ferromagnetism has been reported for 
samples of highly oriented pyrolytic graphite that has been 
irradiated with  high energy protons~\cite{esquinazi,esq2,esq3}. 
These observations have fuelled interest
in the magnetic properties of carbon-only materials,
which are of both great technological and fundamental importance. 
The experimental evidence gathered thus far indicates that the
magnetism does not originate from $d$ or $f$ electrons provided by
chemical impurities but states of $s$-$p$ symmetry unpaired
at structural defects~\cite{esquinazi2}. 
The interest in $s$-$p$
magnetism predates these recent discoveries~\cite{rode,turek,sradnov},
and so does the interest in the technological exploitation of
nanosized particles of graphite, or {\it nanographite}, which
exhibit unusual features depending on their shape,
surface termination and applied field or pressure\cite{toshia}. 
Their tunable and controllable properties
open the way to implementing electro-magnetic devices that are  simultaneously
high-tech, low-cost and easy-to-process.

A graphitic ribbon, created by cutting a graphene sheet along two parallel
lines, exhibits edges, which lower the
dimensionality of the system and, if close enough to each other,
produce a nanosized material. This system is particularly interesting
from a theoretical point of view because the investigation of its
properties as a function of width allows one to explore nanosize
effects as well as bulk effects and the crossover between the two. 
Moreover, graphitic ribbons are of particular interest from the magnetic point
of view as the presence of zig-zag edges disrupts the diamagnetic delocalisation 
of the graphitic $\pi$-electron system creating an instability which can 
be resolved by stabilisation via electronic spin polarisation. 
Several theoretical studies employing the tight-binding model and
density functional theory (DFT) have been devoted  to graphitic 
ribbons~\cite{fujita,hikihara,kusakabe} and
indicate that both charge localisation and paramagnetism
occur in ribbons with zig-zag edges~\cite{waka}.
Experimental evidence of a room temperature paramagnetic response
exists for activated carbon fibers consisting of 
disordered  networks of minute 
graphitic fragments about 2-3 nm across~\cite{enoki}.

Room temperature ferromagnetism requires that long range order exists amongst
the localised spin moments, of such a strength as to survive significant thermal
fluctuations.
While the creation of localised magnetic moments
in these systems has been widely investigated~\cite{arcon,yamashiro}, 
the mechanism, strength and range of the interaction between these
moments are still largely unexplored.
The current  work  addresses the strength and range of the magnetic interactions
amongst the magnetic centres in graphene ribbons with parallel,
zig-zag edges. This is achieved by computing the magnetic interaction
strength as a function of the ribbon's width. 
The method of choice for this work is
hybrid exchange density functional in the B3LYP form.~\cite{bec88,bec93,lyp88}
The mixing of non-local and semi-local exchange present in 
hybrid-exchange functionals, overcomes the major flaws of the local density 
(LDA) and generalised gradient (GGA) approximations in
the prediction of the correct electronic and magnetic ground state in strongly
correlated electron systems~\cite{picket89,Mackrodt93}. 
Furthermore, they   
yield an improved quantitative description of thermochemistry~\cite{ern99}, 
optical band gaps~\cite{mus01}, 
magnetic moments and coupling 
constants~\cite{mar97,bre00,mor02,per01,feng_nio_J_04,feng_cacuo2_04}
 and metal insulator
transitions~\cite{feng_nio04} in strongly interacting systems. This approach
 has also been used succesfully
in recent studies of magnetic ordering in wide band gap
semiconductors~\cite{schmidt_mgti2o4_04}, Prussian blue
analogs~\cite{chan2}, and 
and in fullerene based
metal-free systems~\cite{chan2, chan}. It appears that the hybrid exchange approach allows the
 balance between electron localisation and delocalisation, 
which is crucial in the current work, to be described more reliably than in the LDA and GGA.
The authors, however, are not aware of a systematic study of the perfomances of B3LYP for
systems containing localised (in the sense of non-bonding/atomic) 
orbitals of only {\it s-p} nature. 
The sensitivity of the calculated results to the treatment of electronic exchange 
and correlation is therefore explored by comparison with LSDA and GGA calculations.
In addition to the magnetic stabilisation, alternative mechanisms involving geometric
distortion and charge polarisation are also addressed.

The organisation of this article is as follows: in section II the details of the
system and of the computations are described. In section III, the
electronic structure of the non-magnetic state and 
the eventuality of an instability with respect to
 the electron-lattice interaction are discussed. In section IV
the possible magnetic states (para-, ferro-, and
antiferromagnetic) and the strength of the exchange interaction
as a function of the ribbon width are investigated. Before concluding
in Section VI, section V presents a preliminary study of the likelihood 
of a charge-polarised state as opposed to the spin-polarised one.
 
\section{System's geometry and 
computational details}
\label{det}

\begin{center}
\begin{figure}[!h]
\includegraphics[scale=0.4]{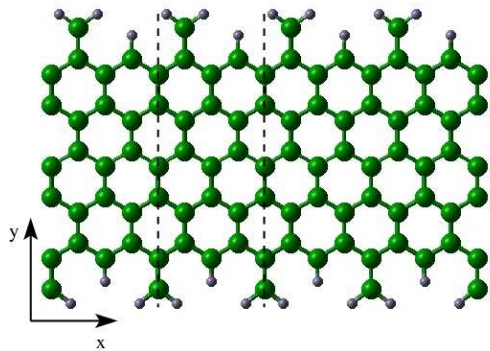}
\caption{(color online) The $N$=5 ribbon with its zigzag edges terminated
by an alternation of methylene groups and H atoms. The dashed
  lines represent the unit cell.}
\label{ucmeth}
\end{figure}
\end{center}

A graphene ribbon of the type shown in Fig.~\ref{ribstr} 
is obtained by cutting a
graphene sheet along two parallel zig-zag lines. 
The ribbon is periodic in the $x$ direction only, and the unit cell is delimited by
the dashed lines. The width of the ribbon along the non periodic
dimension $y$, is defined here by the number, $N$, of {\em
  trans--}polyacetylene-like rows of carbon atoms that run along $x$.  
Concerning the electronic structure of a ribbon, this work 
mainly investigates one type of termination for the dangling
$\sigma$ bonds at the edges: the  mono-hydrogenated
ribbon (Fig.~\ref{ribstr} illustrates this ribbon structure for $N$=5).
Concerning the magnetic properties, the study is extended to the
cases of non-terminated and  methylene group
terminated ribbons, the latter
 being depicted in Fig.~\ref{ucmeth} ($N$=5), 
where the dangling $\sigma$ bonds are saturated alternatively by a
methylene group and a H atom (saturation with methylene groups only in a
planar configuration is prohibited by the steric hindrance
between adjacent methylene groups). 
In the monohydrogenated ribbon, the $\pi$ electron states of the edge atoms
are bound to appear closer to the Fermi level than the $\sigma$ states, 
since the latter provide a stronger bonding interaction than the former.
In the following the focus will be on these highest occupied states.

\begin{center}
\vspace{.2cm}
\begin{figure}[!h]
\includegraphics[scale=0.45]{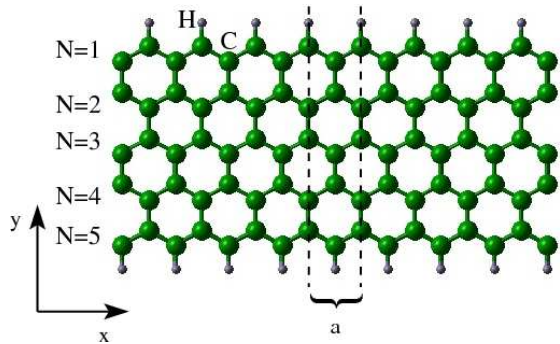}
\caption{(color online) A mono-hydrogenated ribbon of width $N$=5 along $y$. The
  system is periodic only along $x$ and the dashed lines delimit the periodic
  unit cell of length a.}
\label{ribstr}
\end{figure}
\end{center}

The first principles calculations presented here have been performed using
the hybrid exchange density functional B3LYP
as implemented in the CRYSTAL package~\cite{crystal}.
In CRYSTAL, the crystalline wavefunctions
are expanded as a linear combination of atom centred Gaussian orbitals
(LCAO) with $s$, $p$, $d$, or $f$ symmetry. The calculations reported
here are all-electron, i.e., with no shape approximation to the ionic potential or
electron charge density.\\
The geometry optimisations are performed 
  using the alghorithm proposed by Schlegel {\it
  et al} \cite{sch82}. 
The starting geometry was determined by cutting a sheet of graphene
optimised with B3LYP and saturating the dangling bonds with H atoms,
with a C--H distance equal to that of B3LYP-optimised benzene.
The optimisation with B3LYP for graphene gave a lattice constant value
of 2.460~\AA, which is only 0.04\%  smaller than
the experimental lattice constant of natural graphite 
and corresponds to C--C bond lengths of 1.420~\AA. 
The B3LYP optimised C--H bond length of benzene is 1.093~\AA, consistent
with the experimental value of 1.080~\AA~\cite{benzene}. \\ 
Basis sets of double valence quality (6-21G$^{*}$ for C 
and 6-31G$^{*}$ for H) are used.
A reciprocal space sampling on a Monkhorst-Pack grid of shrinking 
factor equal to 60  is adopted after finding it 
to be sufficient to converge the total energy to within $10^{-4}$ eV  per unit cell. 
The Gaussian overlap criteria which control the 
truncation of the Coulomb and exchange series in direct space
are set to $10^{-7}, 10^{-7}, 10^{-7}, 10^{-7}$, and $10^{-14}$.
Typically linear mixing of 70\% and an Anderson second order
mixing scheme is used to guide the 
convergence of the SCF procedure.
Level shifting (typically by  0.3~a.u.) is sometimes required 
to converge the non-spin polarised configuration. 
\newline
In order to compare the energies of various magnetically ordered
states, it is necessary to converge stable self-consistent field
solutions for different electronic spin configurations. This is
achieved by preparing a superposition of aligned or antialigned 
spin densities for particular atomic states to provide 
a suitable initial wavefunction.
It is important to note that that this only affects the initial
wavefunctions and that all solutions presented are unconstrained 
and self-consistent.

\section{Non-magnetic state}
\label{nm}

\subsection{Electronic band structure and density}
\label{nm-ele}

In this section, the properties of the non-magnetic state of a
mono-hydrogenated ribbon of width $N$=10 are discussed. Ribbons with
the other terminations considered here and of different width show a
qualitatively similar behaviour and will therefore not be discussed in
detail. 
Fig.~\ref{NSbands} shows the band structure of (a) the ribbon  
compared to that of (b) an infinite graphene sheet projected onto the
ribbon's first Brillouin zone.
In both cases, only the bands due to the p$_z$ orbitals 
are shown because these states, which sandwich the Fermi level, $E_F$, are the
relevant states for this study.
  The  most striking difference between (a) and (b) affects the highest
  occupied and the lowest unoccupied band, represented with dashed
  lines. The shape of these two bands in (a) and (b) is similar between
  $k=0$ and $k=2\pi/3a$ but becomes markedly different between $k=2\pi/3a$
  and $k=\pi/a$ ($M$). 
In the case of graphene, (b), the two bands become degenerate at $E_F$ at a single point,
  which is equivalent to the K point (point at which the semimetallic
  zero-gap appears) in the first Brillouin zone of
  graphene, and then diverge from each other between K and $k=\pi/a$. In the
  case of the ribbon, the two bands converge until they become
  degenerate exactly at $E_F$ (and the $k$ point at which this happens depends on the
  ribbon's width)~\cite{graphband}, and they remain degenerate and totally
  dispersionless until $k=M$. 
  This results in a high density of states at $E_F$ and this
  electronic structure suggests an instability that might be resolved by
  a number of mechanisms including: (i) a geometric distorsion; (ii) electronic spin
  polarisation; (iii) electronic charge polarisation. The likelyhood
  of each of these eventualities will be investigated in this work.

\begin{center}
\begin{figure}[!h]
\includegraphics[scale=0.35]{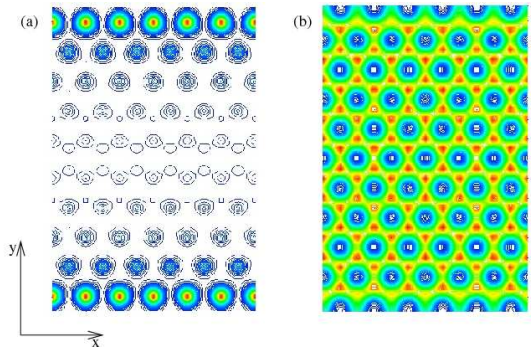}
\caption{(color online) Electron density of a non-magnetic,
  monohydrogenated, $N$=10 ribbon contributed by (a) the states near the Fermi
  level and (b) the rest of the occupied $\pi$ states. 
The range of isovalues is [0:0.028]$e/$\AA$^3$ in case (a)  and  [0:0.12]$e/$\AA$^3$ in case (b).}
\label{eldens}
\end{figure}
\end{center}

\begin{center}
\vspace{.2cm}
\begin{figure}[!h]
\includegraphics[scale=0.5]{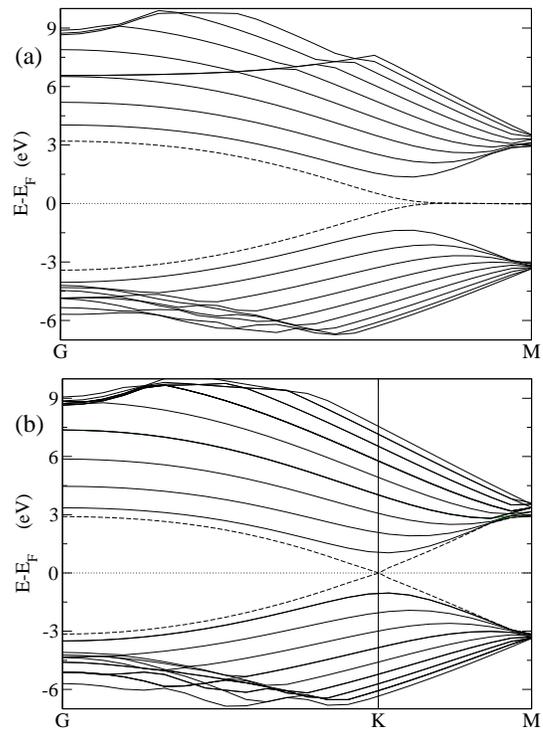}
\caption{ $\pi$ bands in the electronic structure of (a)  a
  non-magnetic, $N$=10, mono-hydrogenated ribbon and (b) a graphene
  sheet representated in the unit cell of the ribbon. Dashed lines
  represent ``edge'' bands and the dotted line is at the Fermi level.}
\label{NSbands}
\end{figure}
\end{center}
By plotting the electron density from states with an energy falling
in a small interval below the Fermi level, the location of  
the dispersionless states at $E_F$ within the ribbon can be visualised.
By choosing  an energy interval of 0.4 eV, the charge density map in
Fig.~\ref{eldens}(a) is obtained.
The electron density is localised mostly at the edges, with the
peaks corresponding to the location of the C atoms at the edges, and
decays into the bulk of the ribbon. In the following, therefore, the two dashed
bands will be referred to as the ``edge'' bands.
Notably, due to the topology of the lattice,
the atoms of the two edges belong to different sublattices of the bipartite
graphene lattice and therefore the degenerate orbitals at E$_F$
 interpenetrate in a non-bonding network. 
Within a tight binding model, Fujita~{\em et al.}~\cite{fujita}  
have shown that non-bonding orbitals alone form the dispersionless part of these bands.
At $k=M$ the states are fully localized at the edges 
and as $k$ approaches $\Gamma$ they  penetrate  into the bulk.

In contrast, projection onto the rest of the occupied bands 
in Fig.~\ref{NSbands}(a) produces the charge density plot 
in Fig.~\ref{eldens}(b).  
The electronic charge contributed by these states is distributed 
across the whole ribbon, and does not decay. The highest values of
this density correspond to the positions of the C atoms with the
exception of the edge's atoms and is significant also along
the C-C bonds, showing a strong bonding character. These states are in nature
the same as the $\pi$ bonding states in graphene and in the following
they will be referred to as ``bulk'' bands.

\subsection{Geometric Distortion}
\label{nm-geom}

As mentioned in the previous section, the high density of states
at the Fermi level present in the non-magnetic ribbons may
indicate an instability with respect to a geometric distortion, which
in this case would be a Peierls distortion as the system is
one-dimensionally periodic~\cite{peierls}.
The prototype system for a Peierls distortion is trans-polyacetylene which, 
interestingly, coincides with the $N$=1 mono-hydrogenated ribbon.
It is therefore interesting to investigate
whether also ribbons with $N$$\ge$2 are unstable with respect to dimerisation.

\begin{center}
\begin{figure}[!h]
\includegraphics[scale=0.35]{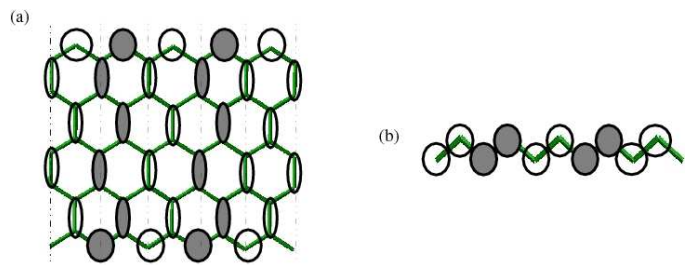}
\caption{(color online) Wavefunction symmetry at the $k=\pi/a$ point 
for a ribbon with $N$=5 (a) and for trans-polyacetylene (b). The
circles represent non-bonding orbitals 
at the Fermi level while the ovals depict those bulk states 
which are about 3 eV below the Fermi level (see
 Fig.~\ref{NSbands}(a)).
The phase of the orbitals is represented by filled ($+$) and empty ($-$) symbols.}
\label{peierls}
\end{figure}
\end{center}

The nature of the electronic states at $E_F$ in
 polyacetylene, however, is drastically different from those
 of a ribbon with $N$$\ge$2.~\cite{kertesz}
By inspection of Fig.~\ref{NSbands}(a) all ``bulk'' bands appear to be 
 nearly degenerate at  $k=M$ (about 3 eV  below the Fermi level). 
By applying a  simple LCAO scheme to a ribbon  with $N=5$ one obtains the 
description displayed in  Fig.~\ref{peierls}(a), where
a wavefunction of those degenerate ``bulk'' bands is superimposed upon
a wavefunction of the two-fold degenerate state at the Fermi level at $M$.
Analogously, for the case of  undistorted polyacetylene, Fig.~\ref{peierls}(b) 
represents one of the possible wavefunctions
 of the two-fold degenerate state at $E_F$. 
In polyacetylene, the non-bonding orbitals are located on nearest neighbours and 
 thus may gain electronic stabilisation energy 
 by forming dimers with a consequent opening of a gap  at the Fermi level. 
 In  the ribbon, the same orbitals are separated by sites in which
orbitals participate in ``bulk'' states. 
In order for the edge states to dimerise
the bulk bonding must be disturbed causing an energy loss 
which competes with the  energy gained from dimerisation. The energy cost
of dimerisation increases with ribbon thickness and therefore one expects
dimerisation to be suppressed at some critical ribbon width.
Our fully relaxed, first principles calculations, predict a bond alternation length 
along the polyacetylene-like chain at the edge is 0.027 \AA~for N=2 and 
smaller than 0.001 \AA~for N=3.
The sharp decrease of the magnitude of the dimerisation with N is a measure 
of the rigidity of the underlying $\sigma-\pi$ bonding structure.

Full structural relaxation of the $N$=10, mono-hydrogenated ribbon
confirms the absence of a dimerisation distortion but at the same time 
a lattice distortion  occurs due to an edge relaxation  effect. 
To describe the topology of the relaxation
we adopt the view in which the ribbon is made up of polyacetylene-like chains
linked by rungs. A carbon atom in the bulk has two intrachain bonds and one interchain 
(rung) bond with its nearest neighbours; at the edge only intrachain bonds are present.
The edge relaxation consists of a uniform contraction (by 0.02 \AA~ in the case of  
$N$=10) of the intrachain bonds at the edge and a consequent expansion of the 
interchain rungs. This effect decrease sharply moving away from the
edge and is to be ascribed to the accumulation of extra bonding charge 
(with respect to the bulk $\pi$ bonds) at the non-bonding part 
of the $\pi$ orbital at the edge. Contrary to the Peierls distortion in polyacetylene, 
not all the non-bonding charge at the edge  participates in this extra bonding 
due to the lack of a bonding counterpart at the atom involved in an interchain bond. 
Crucially this establishes that the localised non-bonding charge 
persists at the edge even after surface  relaxation,
mantaining its tendency towards a magnetic state, as it will be shown in section~\ref{sp}.

\subsection{Very wide ribbons and the graphene limit}
\label{nm-wide}
\begin{center}
\begin{figure}[!h]
\includegraphics[width=6cm,clip]{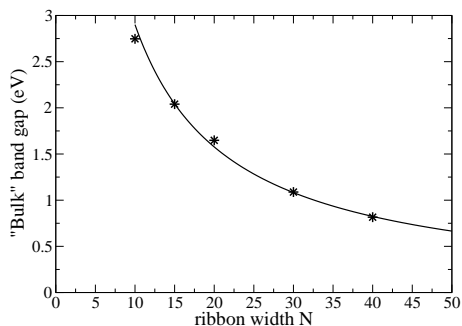}
\caption{The energy gap between the
  highest occupied and lowest unoccupied {\it bulk} states of a
  non-magnetic, mono-hydorgenated ribbon  as a
  function of the ribbon's width.}
\label{bulkgap}
\end{figure}
\end{center}

As the width of the ribbon increases, it is expected that its ``bulk''
  bands, as defined in Sec.~\ref{nm-ele}, will recover the
  graphene limit and the gap between the ``bulk'' occupied and
  ``bulk'' unoccupied states will eventually close. In other words, for
  $N \rightarrow \infty$, the band structure of a
  ribbon will be given by the superposition of the graphene band
  structure and the two ``edge'' bands which, sandwiched between the
  bulk bands, will become
  degenerate,with an energy equal to $E_F$, at $k=\frac{2}{3}
  \frac{\pi}{a}$ (K).
It is therefore interesting to determine the ribbon's width at which
  the band gap between the ``bulk'' states in a ribbon closes. 
By extrapolating the data in Fig.~\ref{bulkgap}, this gap is found to
  recover the graphite limit of a zero band gap, within the room
  temperature thermal energy, when the ribbon width is $\sim 0.25 \mu m$.
This result is independent of whether the geometry of the ribbons is
  constrained at the graphene geometry or fully relaxed.
This critical width is very large, indicating that the screening of
  the perturbation caused by the presence of the edges is very slow
compared to, for instance, the screening of an electronic perturbation at the surface
of semiconducting crystals (typically around 10 nm).
The much larger penetration depth in graphitic ribbons
could be related to the large delocalization energy 
(hopping parameter t $\simeq$ 2.5 eV) typical of carbon aromatic systems, as
  well as to their dimensionality.
Unusually long ranged surface effects 
have been found in previous studies of graphitic systems. In a combined AFM/STM study, 
Ruffieux {\it et al.} observed electronic polarisation
  effects extending over a range of 50$-$60 nm
when H was adsorbed on graphite \cite{ruffieux}.
In a diamagnetic susceptibility
study,~\cite{waka} it was found that the 
the slope of the orbital susceptibility 
as a function of the ribbon width depends on whether 
it has a zig-zag or  armchair geometry.
Surprisingly, the difference between the slope in the two cases
 was observed to increase with the
ribbon's width.

The long-ranged nature of these effects indicates that in the graphene ribbon
a small concentration of defects  would be enough to drastically alter 
the semimetallicity and the conducting properties of the material. 
This conclusion may also be tentatively extended to a wide range of graphitic 
systems but the effects of the three dimensional structure requires further 
investigation.

\section{Electronic Spin Polarisation}
\label{sp}
\subsection{Electronic band structure and density of states}
\label{sp-ele}
As the analysis in Sec.~\ref{nm-geom} excludes the possibility that a
Peierls distortion can resolve the instability brought about by the
high density of states at the Fermi level in the non-magnetic
ribbons, the eventuality of stabilization through spin polarisation is
investigated here.

\begin{center}
\begin{figure}[!h]
\includegraphics[scale=0.6]{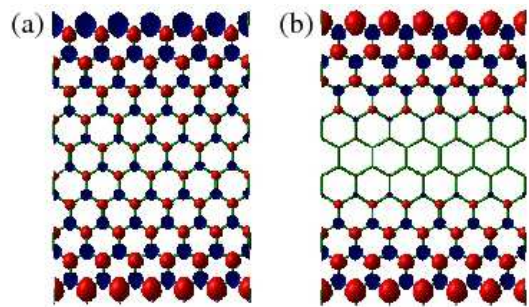}
\caption{(color online) Isovalue surfaces of the spin density for the 
antiferromagnetic case (a) and ferromagnetic case(b). 
The red surfaces represents spin up density and the blue surface spin down density.
The range of isovalues is [-0.28:0.28] $\mu_B/$\AA$^3$ in case (a) and [-0.09:0.28] $\mu_B/$\AA$^3$ in case (b).}  
\label{sd}
\end{figure}
\end{center}

\begin{center}
\begin{figure}
\includegraphics[width=5.5cm,clip]{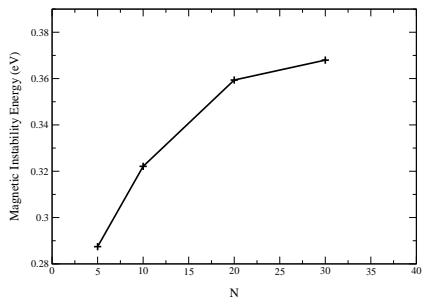}
\caption{The energy difference (per unit cell) between the non-magnetic 
and antiferromagnetic case as a function of N.}
\label{me}
\end{figure}
\end{center}

By allowing the system to be spin-polarised, stable magnetic
states are found, whose spin density
is shown in Fig.~\ref{sd}(a) and Fig.~\ref{sd}(b). Here,
the state depicted in Fig.~\ref{sd}(a) is referred to as 
{\it antiferromagnetic} (AF) because
the spin moments on the C atoms on
one edge are anti-aligned to the spin moments on the opposite edge.
Fig.~\ref{sd}(b) shows the {\it ferromagnetic} (FM) configuration, 
where the spin moments on
both edges point in the same direction. 
Both the AF and FM configurations are found to have a total energy
lower than the
non-magnetic state for all ribbon widths, indicating that spin polarisation is a possible
stabilisation mechanism.
The energy difference between the non-magnetic
and the antiferromagnetic states
plotted in Fig.~\ref{me} is a measure of the strength of the magnetic instability.
The stabilisation is larger for wider ribbons and converges to 
about 0.38 eV per unit cell at $N \sim 30$. Unlike the idealised system presented
here, however, realistic systems are likely to present a mixture of zig-zag and armchair
edges. Since the armchair edges do not give rise to a high density of
states at the Fermi level\cite{dresselhaus}, the magnetic
instability computed here for the idealised system is likely to be overestimated with
respect to the systems experimentally realised thus far. 

The ground state of the system is found to be the AF case, as
  predicted by the spin alternation rule discussed elsewhere \cite{chan} 
and formally expressed in
 Lieb's theorem~\cite{lieb} in the framework of a Hubbard model. It is
  noteworthy that the spin alternation rule is
  satisfied also at the DFT level, which goes beyond the strictly on-site 
treatment of the electron-electron interaction of the Hubbard model.
These results, therefore, extend the validity of the spin alternation
  rule beyond the approximations of the Hubbard model.

\begin{center}
\begin{figure}[!h]
\includegraphics[width=6cm]{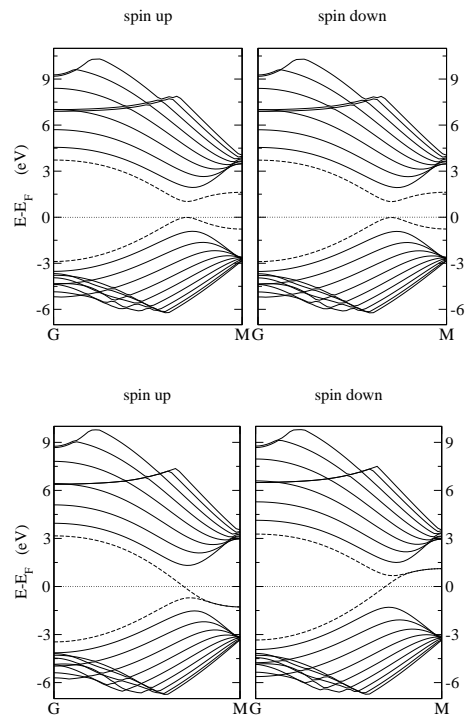}
\caption{Spin polarised band structure of a $N$=10 ribbon. 
The antiferromagnetic case is shown in the upper panels and ferromagnetic
case in the lower panels.
}
\label{bands}
\end{figure}
\end{center}

The analysis of the electronic structure and energetics of these two
stable magnetic states as a function of the ribbon's width
documents the range of the magnetic interaction between the edges and
the nature of the stabilisation mechanism via spin polarisation.
In Fig. \ref{bands} the band structures of the 
  AF (upper panel) and FM (lower panel) states are presented.
The instability of the non-magnetic case discussed in Sec.~\ref{nm-ele} is resolved 
 into two completely different but well-known scenarios: the AF
  solution gives a Slater insulator, whilst the FM solution gives a Stoner metal.

In the AF configuration, opposite spin polarisation on neighboring sites 
localises the electrons in singlet pairing between the two sublattices 
and a band gap opens at $k= \frac{2}{3} \frac{\pi}{a}$,
 where the valence and conduction bands of graphene are
 degenerate (Fig.~\ref{NSbands}). The electronic ground state of a mono-hydrogenated 
 ribbon of finite width is therefore insulating. 
 As the width increases, the band gap  tends to zero 
 following an algebraic decay of the type 1/N
 (as for the ``bulk'' gap of Fig.~\ref{bulkgap}).
The band gap vanishes, within the room
  temperature thermal energy, when N$\simeq$400, 
i.e., for a 80 nm width. This length scale
could be reached by the nanotechonolgy industry in the future
with the synthesis  of single, relatively thin sheets of nanographite.

Conversely, in the case of the metastable FM configuration the
 spin-up and spin-down bands cross in a non-bonding fashion 
at $E_F$ at  $k= \frac{2}{3} \frac{\pi}{a}$. 
The states near $E_F$ are delocalised in the middle of
the ribbon, where no spin density is present. In contrast, 
a small amount of spin
polarisation and electron localisation is present in the same region
in the case of the AF configuration, indicating that the spin
 polarised nature of its ``edge'' states
penetrates deeper into the bulk. The range of ``edge''
states and of the magnetic interactions will be discussed further in
Sec.~\ref{sp-range}.
Notably, no significant difference is detected in the ``bulk'' bands
of the non-magnetic, AF and FM cases,
indicating that magnetic effects on these bands are negligible.
As the ribbon's width increases, the ``bulk'' bands closest to $E_F$ approach each
other  and eventually become degenerate at  
$k= \frac{2}{3} \frac{\pi}{a}$, thus recovering
the graphene band structure.

From a nanotechnological point of view, the AF and FM band structures show 
some appealing features. In fact, the prediction that at the nanometer level 
an AF allignment of the edges' spins produces a strongly insulating state 
and that a FM alignment is conducting, implies that 
one is able to control the conducting properties of a nano-sample by applying
appropriately modulated magnetic fields.

By adopting a mean-field version of the Hubbard model
 with nearest neighbour hopping, it is possible to estimate the values
of the hopping integral, $t$, and of the on-site Coulomb repulsion parameter,
usually referred to as the Hubbard $U$. This is achieved 
 by fitting the computed band structure due to the ``edge'' states at
 $E_F$ with that produced by the Hubbard model.
The resulting values are: $t \simeq 3.2$ eV and $U \sim 2t$. 
Note that this Hubbard $U$ parameter, defined as the direct Coulomb 
integral between the non-bonding localised orbitals (of atomic nature, 
as for the $d$-orbitals in transition metal compounds) 
that exist at the edges of graphene ribbons is 
not to be compared with the $U$ parameter of an
infinite graphene sheet where the $\pi$ orbitals are always delocalised.

It is noteworthy that the values obtained for $t$ and $U$ depend
on the choice of the exchange-correlation functional used
within the DFT calculations. 
Within the gradient-corrected functional PBE~\cite{pbe} one
obtains $t \simeq 2.5$ eV and $U \sim 1.3t$, while
the local density approximation (LSDA)~\cite{vbh}
gives $t$=2.5 eV and $U = 0.9 t$. 

These trends can be understood in terms of the approximation to the 
electronic exchange in the various functionals. The LSDA has a tendency
to delocalise states and suffers from electron self-interaction which, in this
context, results in a lower value for the Hubbard U parameter. The LSDA might therefore
be expected to yield an overestimation of the hopping parameter $t$
and an underestimation of U.
The gradient-corrected PBE functional contains a somewhat better
description of the semilocal exchange which
increases U but does not significantly affect $t$. The B3LYP
functional contains an
element of Fock exchange which explicity compensates for electronic
self-interaction  and
therefore provides a significantly higher U, and tends to localise the
electronic states much more than LSDA or GGA. 
As will be discussed in the next section, a higher degree of
delocalisation increases the strength
of the magnetic interaction but decreases its range.
As discussed above, B3LYP is likely to provide a more reliable description
of this delicate balance but further studies of $s$-$p$ bonded systems are
required to establish this. 

\begin{center}
\begin{figure}
\includegraphics[width=5.5cm]{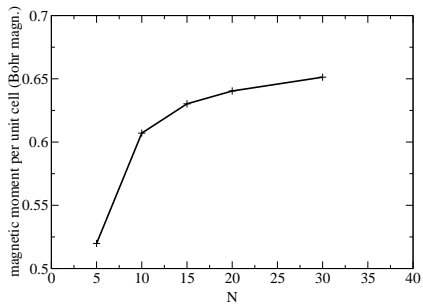}
\caption{The magnetic moment per unit cell as a function of
  the ribbon width in the ferromagnetic state.}
\label{mm}
\end{figure}
\end{center}

Fig.~\ref{mm} shows the magnetic moment per unit cell for 
the FM case which, as expected, approaches the saturation value
of 2/3 $\mu_B$.  As the ribbon widens, the covalent interaction between the 
tails of the edges' wavefunctions diminishes in favor of a purely ionic 
arrangement of the charge which consists in 1/3 $e^{-}$ (for each edge atom)
not participating in any bonding.
 
\subsection{Range of magnetic interaction}
\label{sp-range}
\begin{center}
\begin{figure}[!h]
\includegraphics[width=6cm]{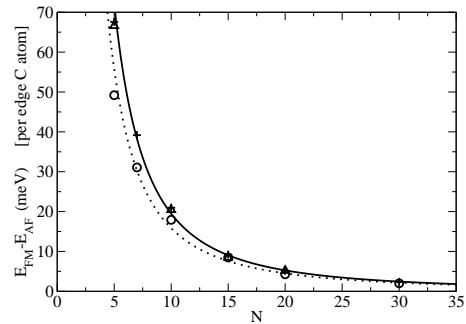}
\caption{The magnetic interaction strength as a function of
  ribbon's width. Empty circles refers to the mono-hydrogenated ribbon
 and are fitted using a power law decay (dotted line) of $1/N^{1.82}$. The
  plus and triangle symbols refer to
 the methylene-group terminated and non terminated ribbons whose decay laws are  $1/N^{1.89}$
 and  $1/N^{1.86}$, respectively.}
\label{magninter}
\end{figure}
\end{center}

As the AF and FM states differ 
by flipping the spin moments
of one edge, the range of the magnetic interaction 
in the ribbons can be measured by computing how the energy difference,
$Delta E$, between
these two states varies as a function of the ribbon's width.
The behaviour of the B3LYP-computed $\Delta E$ as a function of the ribbon's
width is plotted in Fig.~\ref{magninter} for the different types of
termination described in Sec.~\ref{det}. 
The plot refers to unrelaxed ribbons (geometric
relaxation has been found to affect $\Delta E$ only negligibly). 
By fitting the energy variation with N, one obtains a decay law close to $1/N^{2}$ 
for each of the three different terminations considered here: 
the mono-hydrogenated,  the methylene-group terminated 
and the non terminated ribbons. This establishes that the essential
behaviour in these systems
is independent of the particular mechanism used to terminate the zig-zag 
ribbon edge.

Irrespective of the termination
type, a width of $N \sim 8$, corresponding to approximately
14-16 \AA, gives $\Delta E \simeq$25 meV and thus room temperature
magnetic ordering.
This value is significantly higher compared to previous studies based
on standard DFT. 
In a recent LSDA study by Lee {\em et al.}~\cite{lee}, the exchange
interaction is reported to be less than 5 meV at $N=8$. 
Moreover, Wakabayashi {\em et al.}\cite{waka_03} found a value smaller
than 5 meV for $N=3$ in their investigation of the spin wave spectrum
within the random phase approximation of the Hubbard model. 

\begin{center}
\begin{figure}[!h]
\includegraphics[width=6cm,clip]{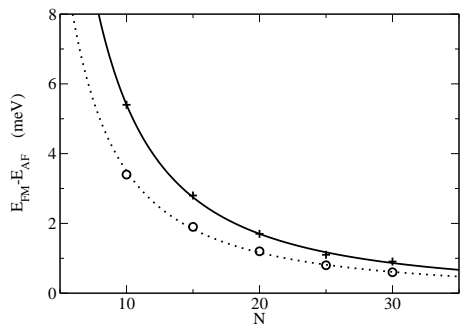}
\caption{Magnetic interaction strength as a function of
  ribbon's width within PBE (plus) and LSDA (circle). The decay power laws are respectively
$1/N^{1.44}$ and  $1/N^{1.39}$.}
\label{deltagga}
\end{figure}
\end{center}

Fig.~\ref{deltagga} displays the same property computed for the
mono-hydrogenated ribbons within the GGA functional PBE 
and the LSDA. 
As noted in Sec.~\ref{sp-ele}, when calculating the Hubbard U the degree of 
charge localisation is smaller in PBE and LSDA than in B3LYP; the magnetic tails of 
the edge states in PBE and LSDA are more delocalised and therefore more long-ranged.
The delocalisation of the state means that the strength of the interaction is lower 
than with B3LYP but its range is increased. The decay power law is $1/N^{1.44}$
within PBE and  $1/N^{1.39}$ within LSDA.
It is interesting to note that only very narrow ribbons could display room temperature
magnetic order according to LSDA and GGA, 
whilst within B3LYP their width is predicted to extend to 1-2 nm.

\section{Electronic Charge Polarisation}
\label{cp}
The instability brought about by the
high density of states at the Fermi level in the non-magnetic
ribbons could, in principle, also be resolved by a charge polarisation mechanism.
When the Hubbard model is extended to include off-site
Coulomb interactions, for instance, the possibility of charge order appears.
Recently, Yamashiro {\em et al.}
investigated the effect of nearest-neighbour interactions for a graphitic ribbon 
adopting an extended Hubbard model 
within the unrestricted Hartree-Fock (UHF) approximation.
In their study it was found that
for some values of the off-site parameter, the charge-polarised (CP)
and spin polarised states compete.\\
The possibility of a CP state is explored here for an $N=10$ ribbon and
within two levels of theory: UHF and B3LYP. 
In the case of UHF, a metastable CP state can be computed in which a 
charge transfer of about 0.05 $e^{-}$
occurs from the C atoms of one edge 
to those of the other. 
The spin-polarised state, however, is much more stable (by 1 eV per cell) than 
the CP state.
In the case of B3LYP, different initial charge densities were tried,
corresponding to charge transfers of 0.05 (as in the stable UHF
solution) and 0.04 $e^{-}$. All such states, however, converge to the spin-polarised
solution. If the proportion of Fock exchange included in the hybrid functional
is increased from 20\% (i.e. the percentage of Fock exchange present
in B3LYP),
a metastable CP state is only achieved at the extreme limit
of 100\%  which corresponds to UHF with a LYP treatment of correlation.
It seems that this extreme limit produces an unphysical instabilty and indeed
it is found that both the  100$\%$ Fock exchange functional and UHF theory 
predict the FM state to be the ground state for larger ribbon widths
which is inconsistent with all other
results and clashes with Lieb's theorem and the spin alternation rule.

\section{Discussion and Conclusion}

In this work, the possible stabilisation mechanisms for a zig-zag ribbon
via geometric distortion, spin polarisation and charge
polarisation have been investigated. 
The influence of the ribbon's width on the
electronic structure and relative stability of the different solutions
were also analysed.

The topology of the non-bonding crystalline orbitals at the edges does
not allow for a significant Peierls distortion for ribbons with N greater than 3 
but rather an edge relaxation  which preserves a strong tendency of the system
to stabilise through a magnetic polarisation, where magnetic moments are localised
at the edges. Two magnetic states were investigated: (i) an
antiferromagnetic configuration where the spin moments on the C atoms
at one edge are antialigned to those on the other edge; (ii) a
ferromagnetic configuration where all spin moments on the C atoms at the edges
point in the same direction. Both states were found to be lower in
energy than the non-magnetic state for any width of the ribbons.
The antiferromagnetic state was found to be the ground state
in agreement with the spin alternation rule and extending Lieb's theorem 
beyond the nearest neighbour approximation.

The magnetic state was found to be about 0.38 eV per unit cell more stable than the diamagnetic 
one for wide ribbons. 
In ribbons that have been realised thus far there is a mixture
of zig-zag and magnetically inactive edges which weaken the tendency 
towards a magnetic state. This is a possible explanation of the recent
scanning tunneling spectroscopy  experiment in which Niimi {\em et al.} \cite{niimi}
measured a peak in the local density of states (LDOS) of edge atoms
signalling the existence of a non-magnetic metallic state (Fig~\ref{NSbands}). 
In fact, if the system were magnetic,
the LDOS at the edge would have shown two symmetric peaks around the Fermi level
representing the opening of a Hubbard gap due to the magnetic polarisation
of the $\pi$ system, as discussed in session~\ref{sp}. 

By fitting the antiferromagnetic band structure to a nearest neighbour Hubbard model 
the value of the $U$ parameter associated with the localised
states at the edges was found to be dependent on the exchange-correlation functional used.
Going from  GGA and LSDA to B3LYP  the $U$ value increases due to 
the explicit inclusion of a self-interaction correcting term in the DFT Hamiltonian.
The self-interaction error is known to be the main cause of the failure
of the LSDA and GGA in the treatment of systems containing localised $d$-orbitals.
However it is not clear at present which treatment of exchange and correlation
provides the most reliable approach for
a system containing localised/atomic-like $s-p$ orbitals.

The magnetic interaction strength was computed as a function of the ribbon width
and, within the B3LYP functional, it was found that ribbons less than 1.5-2 nm
across retain 
magnetic order above room temperature. Within LSDA and GGA the magnetic interaction
strength is weaker but longer ranged.
These three functionals provide differing descriptions of the 
interplay between localisation and delocalisation and its effect on the 
strength and range of the magnetic interaction.
This is evident in the predicted critical ribbon width at which the magnetic
state becomes thermally unstable at room temperature; B3LYP predicts a
critical width of 1-2 nm while LSDA and GGA predict a width an order of
magnitude smaller. 

In conclusion, first principles calculations have been used to establish that the electronic structure of
graphene ribbons with zig-zag edges is unstable with respect to a magnetic polarisation of the edge states.
This mechanism is stable with respect to alternative stablisation routes involving charge polarisation or structural 
distortions.
The magnetic interaction of the edge states is remarkably long ranged.
A band gap in the bulk of the ribbon exists in the antiferromagnetically alligned ground state
with the size of the gap also displaying a very long ranged dependence on the ribbon width closing, at room
room temperature for ribbon widths exceeding 80nm. The qualitative features of these results are robust with
respect to differing treatments of electronic exchange and correlation.
This explicit demonstration of the long ranged nature of the magnetic interaction mediated by the delocalised $\pi$ orbitals
in graphitic systems is consistent with recent observations of magnetism induced by low defect concentrations
in a number of materials.
These results have direct implications for the
control of the spin dependent conductance in graphitic nano-ribbons using
suitably modulated magnetic fields.

\section*{Acknowledgments}

This work is supported by the European Union under the NEST FERROCARBON
project (CEC 012881). The authors would like to thank  S. Bennington and T. E. Weller
for illuminating discussions and the Computational Science and Engineering
Department of the CCLRC for providing the computing facilities.



\end{document}